# Optical signal recording of cellular activity in optogenetic stimulation of human pulp dental cells using a twin-core fiber-based Mach-Zehnder interferometer biosensor

*Faezeh Akbari[1], Mohammad Ismail Zibaii[2], Sara Chavoshi Nezhad [3], Azam Layeghi[2], Leila Dargahi[3], and Orlando Frazao[4]*

1 Department of Biomedical Engineering, University of Kentucky, Lexington, Kentucky, USA, 40504
2 Laser and Plasma Research Institute, Shahid Beheshti University, Tehran, Iran
3 Neuroscience research Center, Shahid Beheshti University, Tehran, Iran
4 INESC TEC, Institute for Systems and Computer Engineering Technology and Science, Porto, Portugal

This paper introduces an innovative two-core fiber (TCF) optic sensor employing a Mach-Zehnder interferometer (MZI) to monitor the optogenetic response of light-sensitive human dental pulp stem cells (hDPSCs). The in-fiber MZI, formed using a segment of TCF optic, detects refractive index (RI) changes in the surrounding medium. The sensor utilizes the evanescent wave of one core as the sensing arm, necessitating a thin cladding achieved through one-sided chemical etching. This design allows the sensor to detect subtle alterations in the RI of the environment by observing displacements in the interference spectrum. The optogenetic stimulation of light-sensitive cells induces variations in ion concentrations, leading to a corresponding change in refractive index. The fabricated sensor, with a peak sensitivity of 675.74 nm/RIU within the RI range of 1.39-1.43, can detect these changes. A computer simulation validated the sensitivity and optimized fabrication parameters, exhibiting satisfactory agreement with experimental results. Spectrum displacements were recorded for both light-sensitive hDPSCs and regular hDPSCs (as a control test). Results from the experiment, analyzed and compared using data analysis software, revealed that 473 nm blue light effectively stimulated light-sensitive hDPSCs. Notably, the proposed sensor, a novel structure, demonstrated its capability to detect RI changes in the cell medium during optogenetic applications.

1. Introduction

In recent years, significant efforts have been dedicated to advancing tools in neuroscience research for evaluating neural function in neurological diseases. Electrophysiological methods stand out as a pivotal approach for measuring neuronal function, identifying pathological functional abnormalities, and analyzing changes in spontaneous and evoked electrical signals in response to therapeutics or disease modeling [1-3]. While electrophysiology is widely applicable, it possesses inherent limitations, notably the absence of cell type specificity and susceptibility to artifacts. Optical technologies, particularly in the realm of optogenetics, have emerged as efficient and reversible means to modulate neuronal activity. These methods enable precise targeting of specific cells or even cell compartments simultaneously, offering high spatiotemporal precision, remote functionality, and non-invasiveness [3]. Optogenetics, a technique in which photons are substituted with electrons for neural stimulation, involves several key steps, as depicted schematically in Fig. 1 (c). These steps include DNA extraction, genome packaging, cell injection, implantation of fiber optics to deliver light to the target area, neural recording, and subsequent data analysis[4]. The application of optical technologies has opened new avenues for treating psychiatric and neurological diseases [5, 6].

Optogenetics serves as a method primarily dedicated to stimulating neural cells, and for monitoring resulting changes, additional techniques or tools are necessary. Table 1 highlights several established optical monitoring techniques widely recognized in the realm of neural imaging. Despite the proven utility and effectiveness of optical imaging modalities in neuroscience, the integration of an all-optical probe capable of both stimulating and recording nerve signals continues to pose a significant challenge. The fundamental basis of neural activities relies on changes in intracellular/extracellular ion concentrations[7], leading to alterations in the refractive index (RI) of the cell medium[8], which is illustrated in Fig. 1. Consequently, fiber optic RI sensors emerge as promising candidates for optically recording neural activity, providing insights into neural behavior.

Table 1. Several established optical monitoring techniques widely recognized in the realm of neural imaging

| Methods | Principle | Applications | Advantages | Disadvantages | Ref |
|---|---|---|---|---|---|
| Two-photon microscopy | Exciting a fluorophore via simultaneous absorption of two photons in a single event | - Neuroscience<br>-Cell biology<br>-Ophthalmology<br>-Developmental biology | - Deep imaging<br>- High-resolution imaging in behaving animal<br>- Optical sectioning | - Complex and massive setup<br>- Requiring ultrashort pulses<br>-Low temporal resolution | [9, 10] |
| Confocal microscopy | Focusing a light source on the sample and using a pinhole to reject out-of-focus fluorescence. | - Neuroscience<br>-Cell biology<br>Imaging fixed samples<br>- Imaging tissues with low sensitivity to photo-damage | - High resolution<br>- Custom designed setup<br>- Optical sectioning<br>-3D imaging | - Limited to use near the tissue surface<br>-Photo-toxicity and photo bleaching | [11] |
| Surface plasmon resonance (SPR) | When the incident light beam strikes some metal surfaces at a particular angle, the SPR results in a graded reduction in the intensity of the reflected light. | -Biosensing<br>-Material science<br>-Neural recording<br>- Measuring RI changes | - Highly sensitive to the refractive index changing<br>-Real time monitoring<br>-label free detection | - Requires a tight contact with target cells due to its shallow penetration depth (405 nm with a 635 nm laser)<br>- Limited to small region<br>- Sensitivity to environmental changes | [12, 13] |
| Optical coherence tomography (OCT) | Low coherence interferometry | - Ophthalmology<br>-Neuroscience<br>-Dermatology | - Noninvasive<br>- Rapid and repeatable<br>- High resolution<br>- In vivo imaging | -Low penetration depth<br>-Complex data analysis<br>-The image quality can be degraded in the presence of media opacity. | [14-16] |
| Near-infrared (NIR) spectroscopy | The NIR light (700–1000 nm) penetrates the skin, the subcutaneous fat/skull, and the underlying muscle/brain, and is either absorbed or scattered within the tissue. | - tissue $O_2$ saturation<br>- changes in the hemoglobin volume<br>- brain/muscle blood flow<br>- Functional activity of the human cerebral cortex | - Noninvasive<br>- Used for human brain research<br>- High temporal resolution | - Limited penetration depth<br>-partial volume effect of scalp and skull<br>-Calibration is required | [17-19] |
| Laser Doppler flowmetry (LDF) | Doppler shift occurs when coherent light is scattered by the moving red blood | - Measuring cerebral blood flow<br>- Measuring blood flow in various organs | - Not requiring a tight contact with the target tissue<br>- Continuous and real-time<br>-Noninvasive | - Sensitive to motion artifacts<br>-Low penetration depth<br>-Calibration challenges | [20, 21] |
| Calcium imaging | Monitoring the activity of the neural circuit using fluorescent calcium indicators | - In vitro and in vivo neural recording Cell biology | - Easy-to-use chemical dyes - High signal-to-noise ratio -Real time monitoring | - Phototoxicity -invasiveness. - Time-consuming -Labeling process challenges | [22, 23] |
| Voltage-sensitive dye (VSD) imaging | Translation of the membrane potential into an optical signal | - Neuroscience -Cardiology -Drug screening | -Real time monitoring -High spatial resolution -Versatility | - No single-cell resolution -Invasiveness - Photo-toxicity - Generation of disruptive oxygen free radicals | [24, 25] |

Table 2. TCF for sensing

| Properties of TCF | Sensing mechanism | Sensitivity/ resolution | Sensing parameter | Sensing range | ref |
|---|---|---|---|---|---|
| A central core and an off-axis core with the same RI | Coupling between the two cores | 3,119 nm/RIU | RI | 1.3160 to 1.3943 | [29] |
| Two similar off-axis cores | Michelson interferometer | 0.0178 per micrometer | Deflection | 0 to 15 micrometers | [30] |
| A central core and an off-axis core with the same RI | Coupling between the two cores | 14.0086 12.0484 nm/(Mol/L) | Salinity measurements | 0 to 5 Mol/L 0 to 1 Mol/L | [31] |
| A two-core photonic crystal fiber | Doppler difference velocimetry | | Velocimetry of particles | | [32] |
| A two-core transversally chirped micro structured fiber optic | Modal MZI | $3.0 \times 10^2$ RIU$^{-1}$ | RI | RI around 1.42 | [33] |
| A two-core photonic crystal fiber | Resonant coupling between the cores | 8500 nm/RIU | RI | RI around 1.23 | [34] |
| Dissimilar doping TCF | MZI | $(78\pm2)$ pm/$\mu\varepsilon$ $(1380\pm20)$ pm/°C | Strain temperature | 0-950 $\mu\varepsilon$ 45 and 80 °C | [35] |
| A two-core photonic crystal fiber | Coupling between the cores | LOD< $10^{-7}$ | RI | | [36] |
| A plastic TCF | Michelson interferometer | $3 \times 10^{-4}$ m$^{-1}$ 70 nm | Bending curvature displacement | | [37] |
| Two off-center cores | Michelson interferometer | 0.096 a.u/mm/s | Flow velocimetry | 0-6 mm/s | [38] |
| A two-core photonic crystal fiber | Interaction between the modes | 5500 nm/RIU (numerical investigation) | RI | 0-100% salinity | [39] |

Since hDPSCs can differentiate into neurons and glial cells under appropriate neurotrophic factors, they can be a good candidate for recording optogenetic stimulations and studying neural activities [26, 27]. In previous work, we investigated the neurogenic differentiation of hDPSCs following optogenetic stimulation[28]. In this study an in-line etched twin-core optical fiber (TCF) sensor based on a Mach-Zehnder interferometer (MZI) was proposed for the optical recording of optogenetic stimulation of hDPSCs.

Due to the favorable characteristics of TCF sensors such as low cost, compact size, stability, and high sensitivity, they have been vastly utilized to construct fiber optic sensors. Various kinds of TCF sensors and their different applications are summarized in Table 2. Here, a piece of TCF was implemented to fabricate an in-line Mach Zehnder interferometer which can provide highly sensitive measurements. For an optimal design, the sensor was simulated based on mode analysis toolbox of COMSOL Multiphysics. A good agreement was obtained between the theoretical and experimental results.

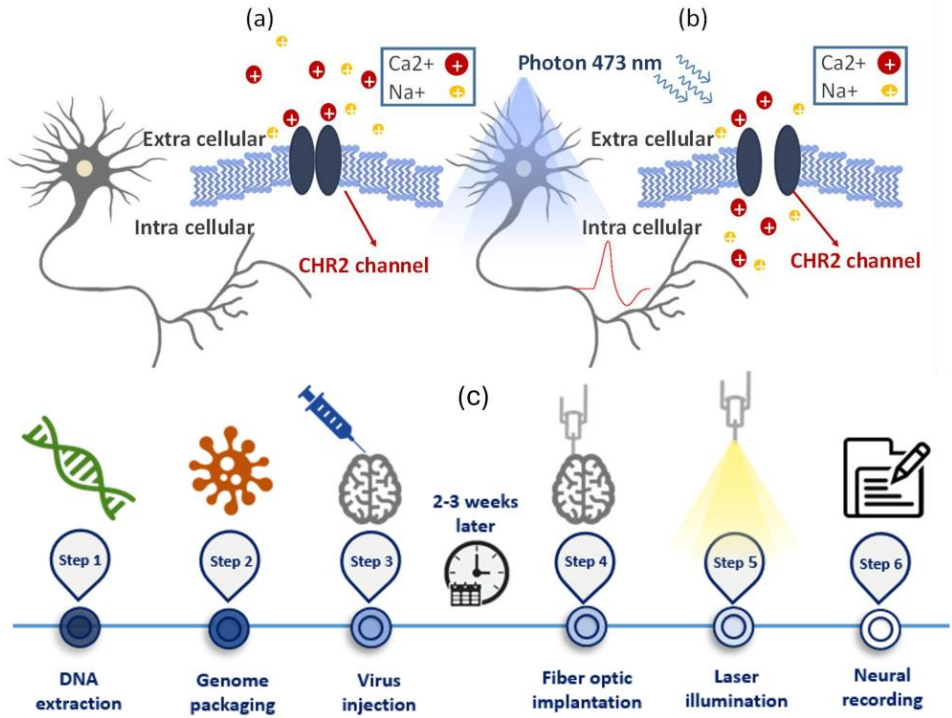

Figure 1. Principles of optogenetic technique. (a) Schematic of the light-sensitive protein ion channel before and (b) after blue light irradiation. Changing the ion densities leads to RI changes in the surrounding medium. (c) The primary steps of the optogenetic technique

## 2. Methods

### 2.1. Theory

The principle of the sensor operation is based on the RI sensitivity of the transmission spectrum of the evanescent wave which can penetrate the ambient medium by reducing the cladding diameter [40-42]. Each core of the TCF acts as an interference arm[43]. By reducing the diameter of the cladding around one of the cores, the evanescent wave of that core penetrates the environment and forms the sensing arm of the interferometer. Changing the RI of the surrounding environment causes a phase change and consequently a wavelength shift in the interference spectrum. The intensity of the light at the output of the interferometer will be given by the following equations [31]

$$I_{OUT}(\omega) = I_1 + I_2 + 2\sqrt{I_1 I_2} \cos\Delta\varphi \qquad (1)$$

$$\Delta\varphi = L \times \Delta k(\omega) \qquad (2)$$

$$\Delta k(\omega) = k_1(\omega) - k_2(\omega) \qquad (3)$$

where L is the length of the interferometer, $I_1$ and $I_2$ represent the intensity of light in the cores, $\Delta\varphi$ indicates the phase difference, $\Delta k(\omega)$ stands for the difference between the wave numbers of the cores, and $\omega$ is the angular frequency. Using the Taylor expansion of $\Delta k(\omega)$ around the arbitrary $\omega 0$, equation (4) is obtained. Hence, the higher order terms of the Taylor expansion can be evaluated:

$$\Delta k(\omega) = \Delta k(\omega_0) + (\omega - \omega_0)\frac{\partial \Delta k}{\partial \omega}\Big|_{\omega_0} + (\omega - \omega_0)^2 \frac{\partial^2 \Delta k}{2\partial \omega^2}\Big|_{\omega_0} + \cdots \tag{4}$$

where the second term is the difference between the inverses of the group velocities of cores 1 and 2 at $\omega_0$ and the third term is the difference between the group delay dispersions (GDD) of cores 1 and 2 at $\omega_0$. Since $\Delta k(\omega)=2\pi(n_1-n_2)/\lambda$, where $n_1$ and $n_2$ are the effective RIs of the cores, the difference between the effective RIs is minute when light propagates in similar media. Therefore, because of their small and negligible values, the higher terms of the Taylor expansion do not contribute to the intensity of the output light which happens in regular interferometers[44]. However, these two terms play a more significant role when light moves through different media such as the dissimilar doped cores of TCFs used to fabricate the proposed sensor. The higher terms of the Taylor expansion led to a quadratic phase function as shown in Fig.2 (a). This figure shows the phase diagram with and without considering the GDD (the higher terms in the Taylor expansion). From now on, the quadratic phase function is referred to as the effect of the GDD in this paper. This quadratic function of the phase can create a special interference spectrum under special conditions. Hence, the interference spectrum of the sensor is different from that of a traditional MZI. As will be explained in the following sections, the interference spectrum shifts oppositely around an arbitrary point in the center (considered 1540 nm). It should be noted that in a very short length of the TCF, the higher order terms of the Taylor expansion can be discarded.

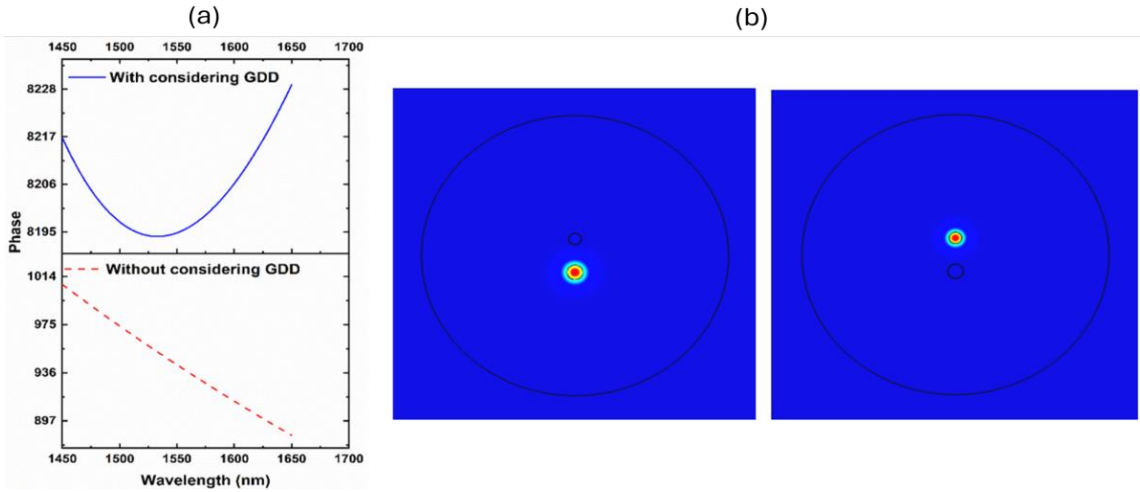

*Figure 2. (a) The effec of GDD on phase diagram. (b) Lack of cross talk between the cores of the TCF*

### 2.2. Simulation of the Sensor

A TCF sensor was simulated in COMSOL Multiphysics to evaluate the sensitivity of the sensor. First, it is necessary to investigate the separate propagation of light in each core and the lack of crosstalk between them along their path. The cladding diameter and TCF length were 125 μm, and 0.4 m, respectively. The host material of the fiber was pure silica. One core is germanium-doped with a diameter of 5.1 μm and a Δn of 0.013, and the other core is phosphorous-doped with a diameter of 6.4 μm and a Δn of 0.00835. The distance between the two cores was approximately 14.9 μm which reduced the crosstalk between them.

Fig. 2 (b) shows the transverse electric field in each core separately where there is no crosstalk between the cores. This model has been assumed to be the cross-section of a compact in-fiber MZI in which light propagates simultaneously through both cores. The input light must enter both fiber cores and be combined in the output to create an interference spectrum. A short piece of coreless fiber (CLF) was used to improve the coupling of light into both cores and reduce the amount of wasted light at the spots where the fibers were spliced. The input light comes from a super luminescent diode (SLD) and the interference spectrum is observed using an optical spectrum analyzer (OSA). A ZEMAX simulation was performed to have an estimation regarding the length of the CLF for an effective coupling. In this simulation, it is assumed that there is another SMF fiber before the CLF, and the numerical aperture that is used in this simulation is equal to the numerical aperture of an SMF (0.1).

Fig. 3(a) shows the ZEMAX simulation results for the effect of changing the length of CLF on the coupling efficiency. The estimated CLF length was obtained about 70 μm. Fig. 3(b) shows the schematic of the sensor. Light from the first CLF coupled to each core of TCF and the second TCF combined the beam which results to MZI interferometric spectrum.

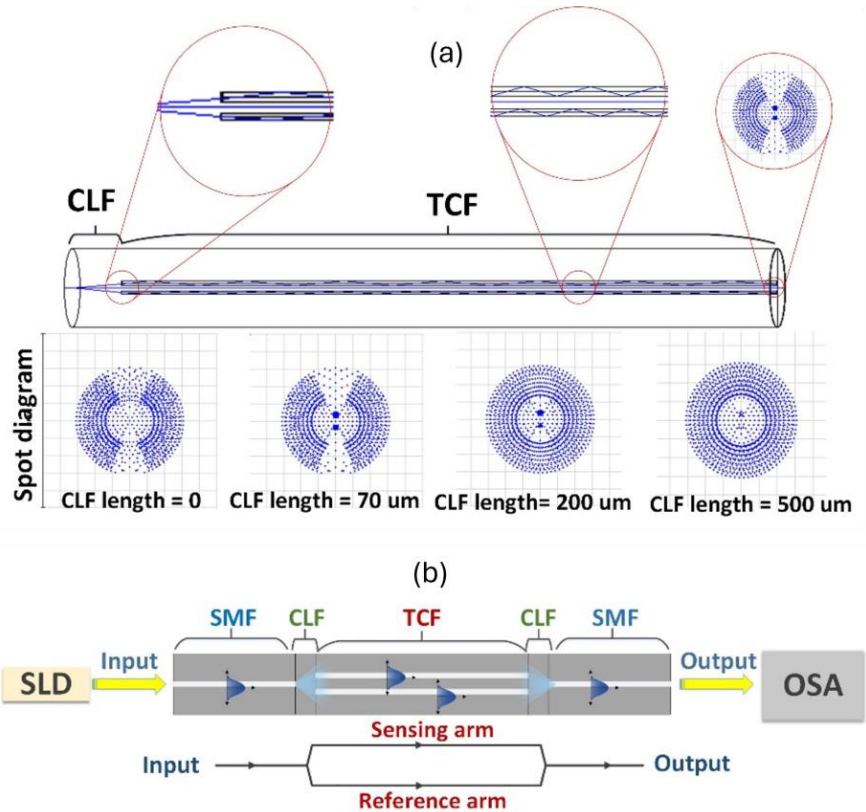

*Figure 3. (a): Simulation results of effective length of CLF, and (b): Schematic of the TCF-MZI sensor*

To enhance the penetration depth of the evanescent wave to the ambient media, it is necessary to reduce the cladding diameter around the sensing cores by using chemical etching. Fig. 5 shows the schematic of the cross-sectional geometry of the TCF sensor after reducing the cladding thickness around one of the cores as a sensing arm and the other core as a reference arm of the interferometer. This cross-sectional shape of the etched fiber is plotted according to our experimental result of one-sided etching on SMF and is used in the simulation. The cladding radius around the sensing core and the length of the etched region were set to be 1 um and 2 cm, respectively.

The RI sensitivity of the sensor in the simulation model was obtained in the range of 1.33 to 1.41 RIU by changing the refractive index of the external medium in the simulation and calculating the effective refractive index of the sensing arm of the MZI. The effective refractive index was used in Equations (1) to (4) to derive to spectrum of the sensor and the displacement of the sensor as a result of changing the refractive index of the external medium.

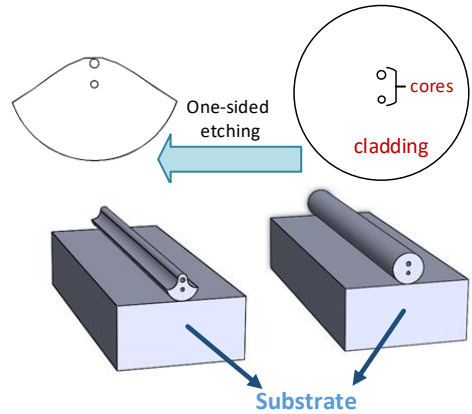

*Figure 5. The cross-sectional geometry of the TCF after one-sided etching. The one-sided etching process which produces this shape will be defined in the following paragraphs.*

## 2.3. Fabrication of the Twin Core Fiber- Mach Zehnder Interferometer (TCF-MZI)

The one-sided etching technique was utilized to reduce the cladding diameter of the TCF. Initially, using a microscope, the direction of the cores was adjusted. Subsequently, the TCF was bonded to a plexiglass substrate with a thin layer of UV-curable epoxy adhesive. Since the cladding around the lower core was glued to the substrate, it was well protected against etching. Conversely, the cladding around the upper core was in direct contact with a Hydrofluoric (HF) Acid Solution. To reduce the cladding thickness, chemical etching was performed for 3 hours using a mixture of 40% HF acid and water (1:2) [20]. Repeating the experiments demonstrated that the cross-sectional shape of the fiber remained relatively consistent after chemical etching. This specific structure of the fiber post one-sided etching can isolate the lower core and enhance the interaction of the sensing core with the surrounding environment. Finally, a 1×2 cm² chamber made of plexiglass was used to accommodate the fiber and hold the medium during the experiments.

The length of the TCF is another critical parameter that requires investigation. The GDD does not significantly contribute to the short length of TCFs. Hence, different lengths can result in entirely distinct interference spectra. Fig. 6 displays the experimental and theoretical results regarding the effect of changing the lengths of the TCF on the interference spectrum across a wide wavelength range. A good agreement is observed between the theoretical and experimental results. According to the findings, the minimum length for the TCF falls within the range of 30-40 cm in order to observe a sufficient number of peaks and dips on both sides of the spectrum. The interference spectrum indicates that the effect of GDD is not observed when the TCF has a short length, as already anticipated. Moreover, when the TCF is too long, it produces a sharp spectrum; however, it becomes unstable for practical experiments.

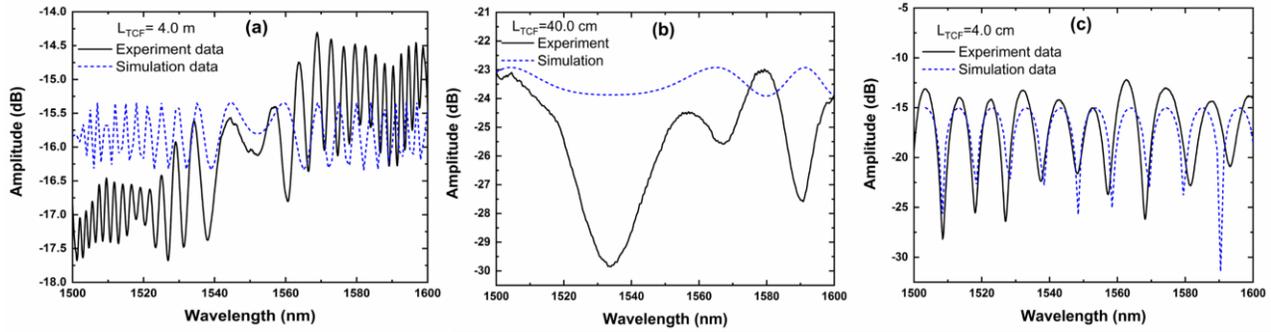

*Figure 6. Comparing the spectra in the experimental and theoretical results of the TCF in different lengths: (a) 4 m, (b) 40 cm, and (c) 4 cm*

## 2.4. Experimental Setup

The experimental setup is schematically depicted in Fig. 7(a). The light source is an SLD with a central wavelength, full width half maximum (FWHM), and power of 1547 nm, 79.6 nm, and 50 μW, respectively. After passing through a dual-stage isolator, the light from the SLD reached the TCF-MZI sensor. The spectral response of the sensor was observed using the OSA. The data from the OSA were recorded by a GPIB interface on a computer and processed using programmed code in the LabVIEW software. The cells were stimulated using a blue-light laser diode (473 nm) with a frequency of 15 Hz, pulse number of 120, duty cycle of 50%, stimulation time of 60 s, and rest time of 300 s. The laser pulses were controlled by a DAQ card and the LabVIEW program. As shown in Fig. 7 (b), Poly-L-lysine (PLL) was used to immobilize the hDPSCs on the fiber sensor [32]. PLL contains positively charged amino groups that can bind to a negatively charged silica surface, facilitating cell adherence to a positively charged PLL-treated TCF surface [31]. For immobilization, the sensor surface was washed with 38% HCl to neutralize the negative charge. Then, it was rinsed with deionized water three times at room temperature. The fiber was dried in a microbiological safety cabinet for thirty minutes. Next, a PLL solution (0.1%) (Sigma, P8920) was added to the chamber and allowed to evaporate overnight. Fig. 7(c) shows the shift resulting from changing external RI. The immobilization of hDPSCs on the sensor surface was detected by recording the wavelength shift of the spectrum over time. According to the Langmuir adsorption model, the wavelength shift, as depicted in Fig. 7(d), indicates that the cells have successfully attached to the sensor surface.

## 2.5. hDPSCs Preparation

In this experiment, hDPSCs isolated from third molars were cultured in DMEM/F12 (1:1) (GIBCO-BRL) supplemented with 10% fetal bovine serum (FBS, Invitrogen), 2 mM L-glutamine (Sigma-Aldrich), and 1% (v/v) penicillin-streptomycin, and incubated at 37 °C with 5% $CO_2$ and humidity conditions of 95%. The cells were monitored daily and the media were changed every 3 days. The cells were sub-cultured once they reached 80-90% confluence. The hDPSCs at the passage 3-4 were seeded in T-25 flasks at a density of $5\times10^5$ cells per flask in the basal medium. After 48 h, the cells were transduced with lentiviral construct encoding hChR2 (H134R)-mCherry. 96 h post-transduction, the mCherry fluorescent signal was observed using an inverted fluorescence microscope (Olympus). The non-opsin-expressing hDPSCs were considered as the control group in this study. As shown in Fig. 10(d), the majority of the cells were red fluorescence-positive as compared to the control cells in Fig. 10(c). After the transduction efficiency was verified, the hDPSCs were detached from the flasks using low trypsin concentrations (0.025%), pelleted by centrifugation, resuspended in serum-free media, and then used for further experiments.

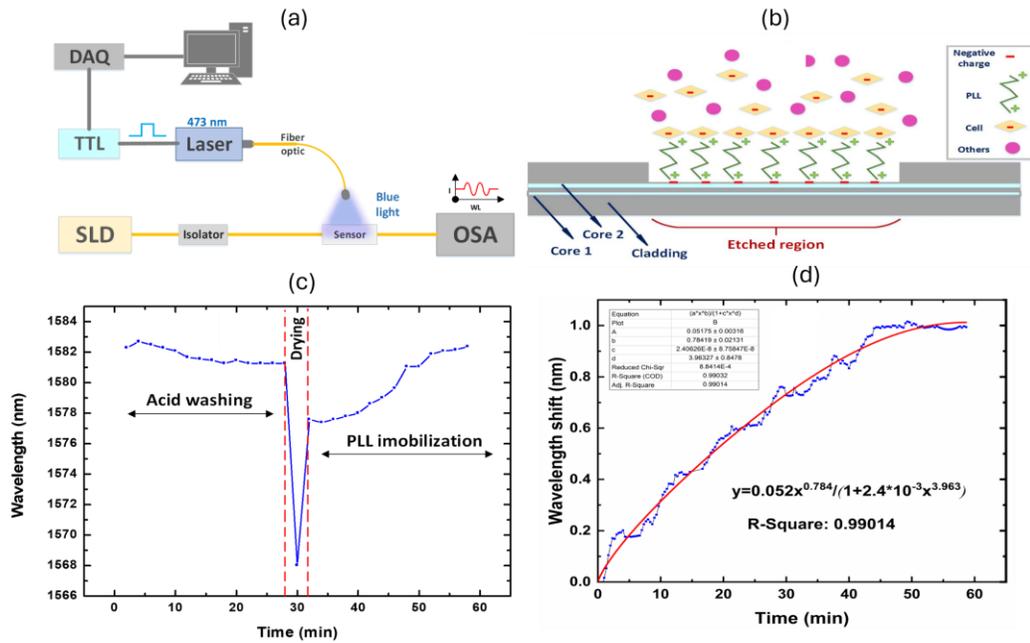

*Figure 7. (a) Schematic of the experimental setup. (b) Schematic of the immobilization of the hDPSCs on the sensor surface using PLL, (c) the wavelength shift of the sensor over the time with acid washing and PLL over time, and (d) the wavelength shift of the cell immobilization verified by the Langmuir adsorption model.*

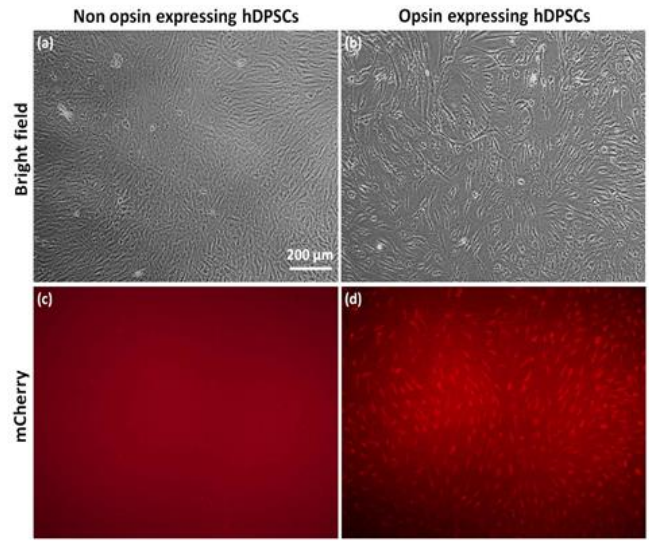

*Figure 8. The expression of the mCherry protein in the hDPSCs 96 hours after transduction. The representative phase-contrast microscopy micrographs of (a) the non-opsin-expressing hDPSCs and (b) the opsin-expressing hDPSCs. The representative fluorescence micrographs of (c) the non-opsin-expressing hDPSCs, and (d) the opsin-expressing hDPSCs. Scale bar=200 µm.*

## 3. Results and Discussions

## 3.1. RI Sensitivity
### 3.1.1. Theoretical Results
Fig. 9 (a) and (b) depict the theoretical results of the wavelength shift at different RIs. The results indicate that the wavelength shifts of the two sides of the spectrum behave differently. This behavior is expected due to a quadratic phase function and the GDD effect. As the RI increases/decreases, the two sides of the interference spectrum move away/towards each other. The sensitivity diagram is plotted for a cladding radius of 1 μm in Fig. 9 (a). Simulation results demonstrate that the RI sensitivity for dip R in the range of 1.33-1.39 RIU and 1.39-1.43 RIU is 326.6 nm/RIU and 685.75 nm/RIU, respectively. Additionally, the RI sensitivity for dip L in the range of 1.33-1.39 RIU and 1.39-1.43 RIU is -292.60 nm/RIU and -568.08 nm/RIU, respectively. To achieve higher sensitivity, the wavelength separation of the two sides (R, L) of the spectrum around the central part was calculated. In this case, the RI sensitivity was determined to be 619.205 nm/RIU and 1253.82 nm/RIU in the range of 1.33-1.39 RIU and 1.39-1.43 RIU, respectively. The simulation results indicate that a narrower cladding diameter around one of the cores can enhance the sensitivity of the sensor.

### 3.1.2. Experimental Results
The sensitivity of the sensor was assessed using a solution of water and glycerin (0-80%) within the RI range of 1.33-1.43 [31]. Fig. 9(c) illustrates the output spectrum of the sensor at different RIs. The obtained results indicate that the RI sensitivity of the R and L dips within the range of 1.33-1.4 RIU is 130.0 nm/RIU and -99.56 nm/RIU, respectively. Within the range of 1.4-1.43 RIU, the RI sensitivity of the R and L dips is 868.51 nm/RIU and -394.43 nm/RIU, respectively. Fig. 9(d) displays the wavelength shift for different RIs, which aligns with the theoretical predictions (Fig. 9(b)). Due to the quadratic phase, both sides of the spectrum behave oppositely. To achieve higher sensitivity, the separation of the two sides of the spectrum around the central part was calculated, as shown in Fig. 9(e). In this scenario, the sensitivity is 233.62 nm/RIU in the range of 1.33-1.4 RIU and 870.01 nm/RIU in the range of 1.4-1.43 RIU, with $R^{^2}=0.99$. These results indicate that the TCF-MZI sensor could be a promising candidate for optical neural recording. The disparity between the theoretical and experimental results arises from imperfections in the one-sided etching process and the inability to measure the thickness of the fiber cladding after one-sided etching.

## 3.2. Detection of Optogenetic Stimulations of hDPSCs Using the Fabricated TCF-MZI
The opsin-expressing hDPSCs were optically stimulated according to the specified stimulation parameters. The results of optical neural recording for optogenetic stimulation are presented in Fig. 10. The blue vertical arrows indicate the laser stimulations, comprising a series of laser pulses with the stimulation duration of 60 s at 473 nm. Fig. 10(a) depicts the situation before and after optical stimulation, which leads to an oscillation in the wavelength of the interference spectrum. Fig. 10(b) illustrates a typical wavelength shift during 58 minutes of optogenetic stimulation. ChRs are non-selective cation channels permeable to $Na^+$, $K^+$, and $Ca^{2+}$. Upon illumination and subsequent opening, they depolarize the membrane. The wavelength shift of the sensor results from integrated RI changes within a small measurement volume, and the RI can be altered by changes in ionic composition [33]. Additionally, in previous work, we demonstrated that optogenetic stimulation of the entorhinal cortex promotes hippocampal neurogenesis and synaptic plasticity. Electrophysiological methods, utilizing single-unit signal recording, revealed a significant increase in the firing rate of the cornu ammonis (CA1) pyramidal neuron relative to the baseline firing after light stimulation of the medial entorhinal cortex [34]. Therefore, the neural signal is associated with changes in ion concentration and effective RI, and the sensor could detect RI changes in the hDPSCs during optical stimulation. As a control test, non-opsin-expressing hDPSCs were immobilized on the sensor. Fig. 10(c) demonstrates that the wavelength shift is solely due to cell immobilization, and illumination does not stimulate the hDPSCs. This confirms that optogenetic stimulation leads to nervous system activation and changes in ionic concentrations.

It's important to note that this is a case study demonstrating the capability of the proposed sensor to detect optogenetic stimulation of light-sensitive cells. In future studies, more samples need to be tested, and statistical analysis is necessary to determine if the detected changes are statistically significant. Fiber optic biosensors, due to their small size and immunity to electromagnetic interference, can replace optrodes in optogenetics techniques for simultaneous optical recording and stimulation. Additionally, optrodes can be utilized for neurotransmitter detection in behavioral neuroscience research.

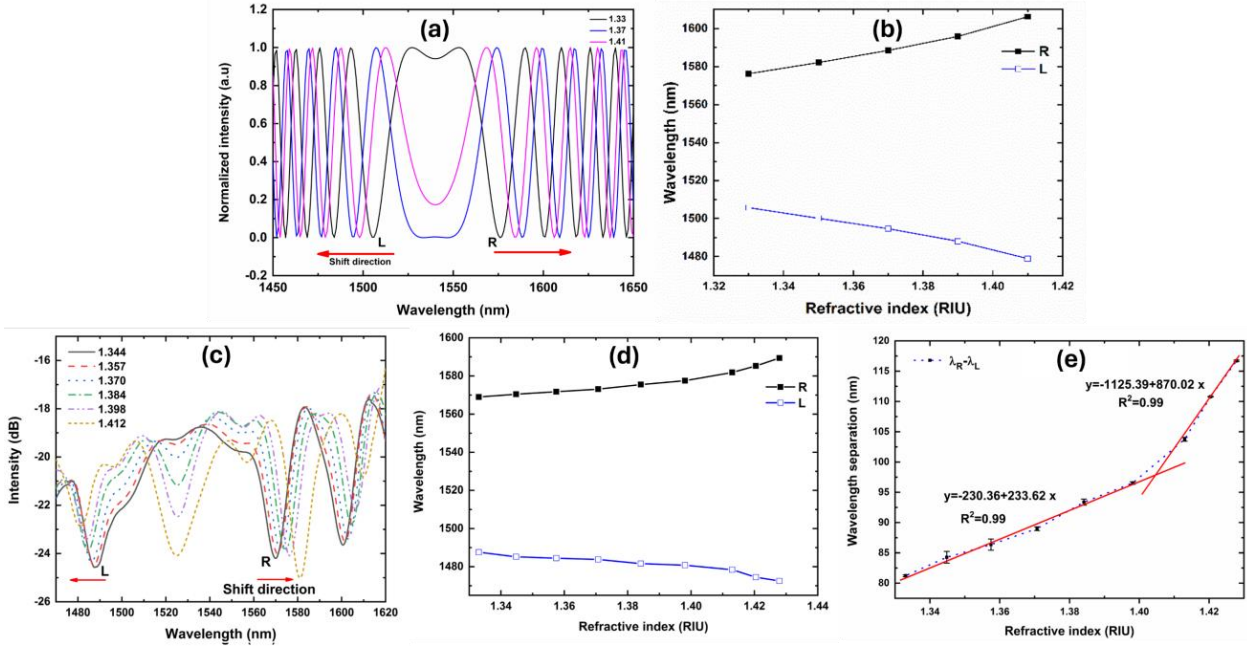

*Figure 9. (a) Theoretical spectrum of the sensor. (b) The wavelength shift of peak R and L versus RI (theoretical results). (c) The output spectrum of the sensor, (d) The wavelength shift of peak R and L versus RI (experimental results), and (e) Wavelength separation (R and L) versus RI.*

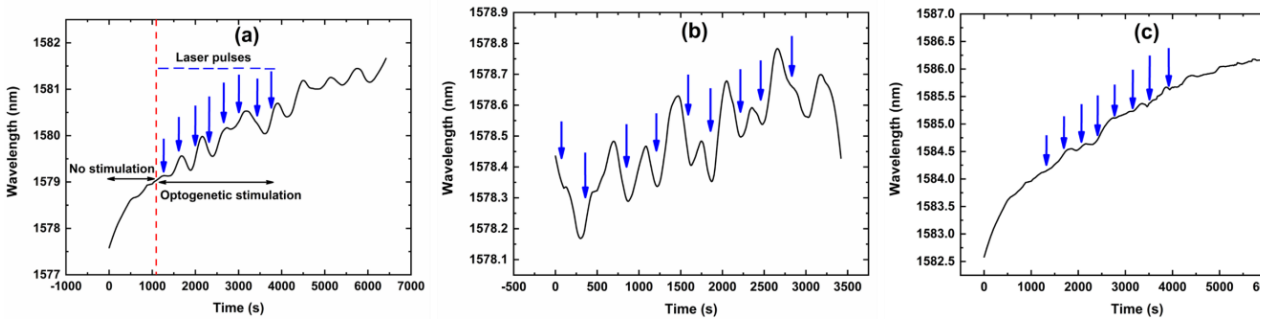

*Figure 10. (a) and (b) The stimulation of the opsin-expressing hDPSCs. (c) The stimulation of the non-opsin-expressing hDPSCs. In the stimulation of the opsin-expressing hDPSCs, the spectrum is recorded in an extended time to find out the behavior of the cells before and after stimulation.*

## 4. Conclusions

A new method for monitoring cell activity using fiber optic RI sensors was reported based on TCF-MZI. To enhance the RI sensitivity of the sensor, one side of the cladding of the TCF was etched. With the GDD effect, the RI sensitivity of the sensor was 233.62 nm/RIU in the range of 1.33-1.4 RIU and 870.01 nm/RIU in the range of 1.4-1.43 RIU with $R^2=0.99$. There was good agreement between the theoretical and experimental results in terms of sensor sensitivity. The obtained results show that the TCF-MZI sensor can be a good candidate for optical neural recording. Therefore, the sensor was used for optical recording of neural activity of optogenetic stimulation of the opsin-expressing hDPSCs. Optical stimulation of opsins led to changes in ion concentrations and consequently, the RI of the extracellular medium. Wavelength shift and spectrum displacement due to changing effective RI of the surrounding medium of the sensor can be a measure of neural activity. The obtained results show that RI fiber optic sensors can be used for optical recording in optogenetic stimulation where an all-optical device is demanded. Since all-optical probes in optogenetic techniques are highly needed, the proposed sensor can be used for optical recording and stimulation simultaneously. Also, it can be integrated with fiber optic biosensors for the detection of biologically significant neurotransmitters as a multifunctional optical probe.

## Acknowledgments

We would like to thank Dr. Karl Deisseroth for their generous gift of the plasmids.